\input{aipcheck}

\documentclass[
    ,final            % use final for the camera ready runs
,numberedheadings % uncomment this option for numbered sections
  ]
  {aipproc}

\layoutstyle{8x11double}

\usepackage{graphicx}% Include figure files
\usepackage{subfig}
\usepackage{epstopdf}
\usepackage{dcolumn}% Align table columns on decimal point
\usepackage{amsmath}

\begin{document}

\title{Radio emission from air showers. Comparison of theoretical approaches.}

\classification{95.85.Ry, 41.60.Bq, 03.50.-z, 41.20.J}
\keywords      {radio emission, moving charge, medium, extensive air shower, cosmic rays}

\author{Konstantin Belov}{
  address={UCLA Department of Physics and Astronomy / 154705, 
475 Portola Plaza, 
Los Angeles, CA 90095, USA}
}

\begin{abstract}
While the fluorescence and the ground counter techniques for the detection of ultra-high energy cosmic rays (UHECR) were being developed for decades, the interest in the radio detection diminished after the initial experiments in the 1960s. As a result, the fluorescence and the surface array techniques are more mature today, providing more reliable measurements of the primary cosmic particle energy, chemical composition and the inelastic cross-section. The advantages of the radio technique are 100\% duty cycle and lower deployment and operational costs. Thus, the radio technique can greatly complement the fluorescence and the ground array detection and can also work independently. With the ANITA balloon detector observing UHECRs and the success of LOPES, CODALEMA and other surface radio detectors, the radio technique received a significant boost in recent years. Reliable Monte Carlo (MC) simulations are needed in order to obtain the energy and other parameters of the primary cosmic ray particle from the radio observations. Several MC techniques, like ZHairesS and the Endpoint Formalism, were proposed in recent years. While they seem to reproduce some of the observed data quite well, there is a divergence between the different approaches under certain conditions. In this work we derive these approaches from Maxwell's equations and prove their identity under certain conditions as well as discuss their applicability to the UHECR air showers and to a proposed experiment at SLAC.
\end{abstract}

\maketitle

%%%%%%%%%%%%%%%%%%%%%%%%%%%%%%%%%%%%%%%%%%%%
%% MAINMATTER
%%%%%%%%%%%%%%%%%%%%%%%%%%%%%%%%%%%%%%%%%%%%

\section{Introduction}

Ultra high energy cosmic rays (UHECRs) are charged particles with the energy above $10^{18}$ that are constantly bombarding the earth. Even after decades of intensive study, the origin, acceleration mechanism and chemical composition of the UHECRs are not yet known. While a correlation between the UHECR arrival directions and the nearby active galactic nuclei has been reported \cite{Abreu2010314}, the correlation becomes weaker as more data is collected. A separate study does not find any correlation \cite{Abbasi2008175}. In addition, a Greizen-Zatsepin-Kuzmin mechanism (GZK), described as a photo disintegration of the nucleus on cosmic microwave background, see \cite{greisen:gzk66}, limits the UHECRs travel distance, effectively excluding the distant sources. A flux suppression that could be interpreted as the GZK cutoff was independently seen by two large cosmic ray observatories \cite{Abbasi200953, Abraham2010239}. Also, current theoretical models, with the exception of the most exotic ``top down'' models, fail to predict mechanisms capable to accelerate the cosmic rays above $10^{20}$ eV, yet the cosmic rays above that energy were observed \cite{PhysRevLett.10.146}. The controversy in the chemical composition of the UHECRs has persisted for a long time \cite{PhysRevLett.104.161101, PhysRevLett.104.091101}. Despite so many unanswered questions, the UHECRs remain to be the only source of elementary particles with energies above the reach of modern accelerators and allow us to study particle interactions at these energies \cite{Belov2006197}. In addition, the UHECR research can be considered as ``non-electromagnetic'' astronomy, complementing research in different electromagnetic frequency bands. 

The cosmic ray spectrum falls rapidly with energy, and the very low cosmic ray flux makes direct observations at the energies above $10^{14}$ eV impractical. In order to increase the detector volume, the earth's atmosphere can be used as a giant calorimeter. Cascades of secondary particles in the atmosphere (extensive air showers) can be detected by air fluorescence or Cherenkov radiation in UV and by particle counters on the ground. The modern hybrid cosmic ray telescopes utilize two or all three of the above mentioned techniques simultaneously \cite{Tameda200974, Abraham200450}. The air fluorescence technique, which is capable of observing a large portion of an extensive air shower in the atmosphere, provides very good primary particle energy resolution combined with a precise pointing to the arrival direction. Detailed information about the shower profile, the number of charged particles as a function of the slant depth in the atmosphere, and the depth of the shower maximum ($X_{max}$) in particular, can be used for chemical composition and particle cross-section study \cite{PhysRevLett.104.161101, PhysRevLett.104.091101, Belov2006197}. The UV observations, however, can only be conducted during moonless nights, which limits the air fluorescence and the air Cherenkov detectors duty cycle to not more than 10\%. The ground particle counters can be operated continuously, increasing the data statistics, but the energy and the shower profile reconstruction for this method relies on extensive Monte Carlo (MC) simulations, and is not yet as accurate as the air fluorescence technique.

Detection of the extensive air showers caused by UHECRs by radio emission was first used in the 1960s \cite{allan1970}. The success however was somewhat limited due to a lack of understanding of the radio emission mechanism. The later efforts were concentrated more on the air fluorescence, air Cherenkov and the ground counter array methods, but the radio observation technique is very appealing with such an advantage as 100\% duty cycle combined with lower deployment and operational costs. The advancements in modern computing for the extensive MC simulations led to a better understanding of the radio emission mechanism and the new ground and balloon-borne experiments are in operation or planned \cite{doi:10.1117/12.551466, Ardouin2005148, 2011arXiv1109.5805C, Fuchs201293, PhysRevLett.105.151101}. 

% Reconstructing the energy and Xmax of the UHECRs from radio relies on MC.
% Xmax is needed for composition and cross-section study.
Similar to a ground counter array, the energy and the $X_{max}$ reconstruction from radio data relies on the MC simulations, which, in turn, require a complete understanding of the emission from a moving charge in a dense medium.  A ``full'' MC approach tracks radio emission from each individual particle and takes into account coherence effects of the air shower for a ground observer \cite{Ludwig2011438, AlvarezMuniz2012325}.  The results of the MC simulations derived using different approaches start to agree with each other, but only in the far zone and away from a low frequency limit \cite{clancy-arena2012}. In this work, we compare the different theoretical approaches to the problem of the emission from a charged particle in a dense medium to study their validity and limitations. 

\section{Emission from a moving charge in a dense medium}

\begin{figure}[htp]
\includegraphics[width=\columnwidth]{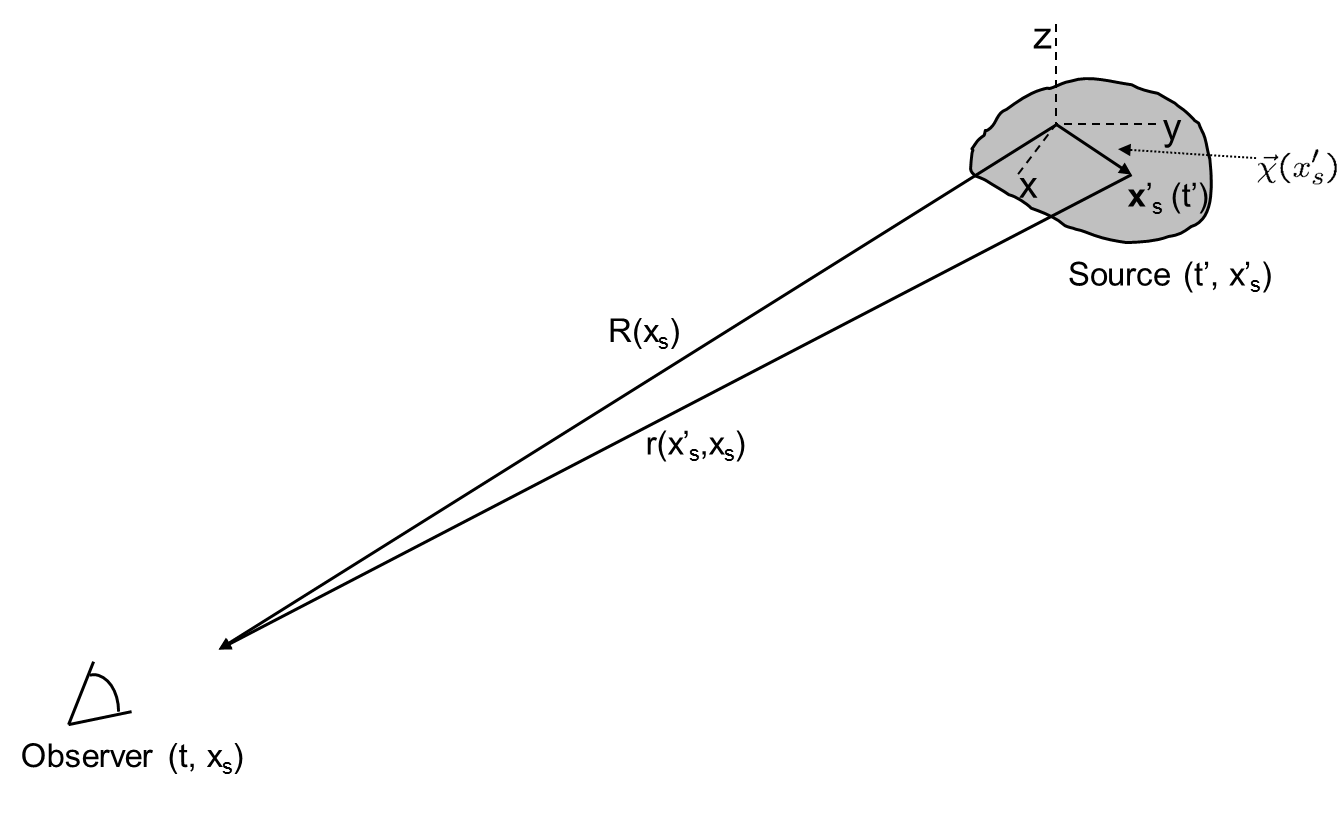}
\caption{Radio emission from a source in a dense medium.}
\label{ChargedDiagram:fig} 
\end{figure}

Despite appearing to be a ``classical'' problem, the electromagnetic emission from a moving charge in a dense medium does not have a complete analytical solution yet. The problem was first formulated by Frank and Tamm in 1937 \cite{franktamm:37}, but was only solved for some special cases such as the far zone approximation, infinite particle tracks, slowly changing properties of the medium, a particle passing through a thin plate and a few others \cite{ginzburg:46}. For an infinite track in vacuum the classical solution can be derived from Maxwell's equations with no difficulty \cite{landaufieldtheory:88, jackson91}.  In case of a finite track in a dense medium with a constant index of refraction $n>1$ the track can be split into small straight segments and then two different approaches can be used:
\begin{itemize}
\item only the emission from the straight part of a segment, where the particle moves uniformly in a dense medium, is taken into account. The emission from the ends of the track is ignored assuming that the emission time is very short \cite{PhysRevD.45.362}.
\item only the emission from the ends of a segment, where the particle accelerates or decelerates, is taken into account, assuming that the particle does not emit while moving uniformly in vacuum \cite{PhysRevE.84.056602}. The medium effects are taken into account at the end by adding the index of refraction.
\end{itemize}
To derive both approaches, let's consider a charged particle moving along the track $\vec{\chi}(x'_s)$, see the diagram on Figure \ref{ChargedDiagram:fig}, where apostrophe indicates a delayed time or coordinate.

\subsection{Summing over the straight segments of the track.}

Let's first express electric and magnetic fields using electric and magnetic potentials:
\begin{align}
&\vec{E}=-\vec{\nabla} \phi - \frac{\partial \vec{A}}{\partial t}\\
&\vec{B}=\vec{\nabla}\times\vec{A}
\end{align}
where the delayed potentials at time $t^{'}=t-\frac{ r ( x_s, {x_{s}}^{'} ) } {c}$ are:
\begin{align}
&\phi(x_s, t)=\frac{1}{4\pi\epsilon_0} \int \! \frac{\rho(x_{s}^{'},t)}{r(x_s,{x_{s}}^{'})} \, \mathrm{d{\bf V'}} \\
&\vec{A}(x_s, t)=\frac{1}{4\pi\epsilon_0 c^2} \int \! \frac{\vec{j}({x_{s}}^{'},t)}{r(x_s,{x_{s}}^{'})} \, \mathrm{d{\bf V'}}
\end{align}
and $\rho(x_{s}^{'},t)$ and $\vec{j}({x_{s}}^{'},t)$ are the charge and current densities respectively, and $x_{s}$ and ${x_{s}}^{'}$ are the particle coordinate in the observer and the source reference frames. 
With the Fourier components of the fields defined as:
\begin{align}
&\vec{E}(x_s,t)=Re \int \! \vec{E_{\omega}}(x_s)e^{-i\omega t} \, \mathrm{d}\omega\\  
&\vec{B}(x_s,t)=Re \int \! \vec{B_{\omega}}(x_s)e^{-i\omega t} \, \mathrm{d}\omega
\end{align}
the electric  and magnetic field spectral components can be written as:
\begin{align}
&\vec{E}_{\omega}(x_s)=\frac{1}{4\pi\epsilon_0} \int \! \rho_{\omega}\nabla'\left(\frac{e^{i\vec{k}\vec{r}}}{r}\right) \mathrm{d}{\bf V'} + \frac{ik}{4\pi\epsilon_0 c} \int \! \vec{j}_{\omega} \frac{e^{i\vec{k}\vec{r}}}{r} \mathrm{d}{\bf V'} \nonumber \\
&\vec{B}_{\omega}(x_s)=\frac{1}{4\pi\epsilon_0 c^2} \int \! \vec{j}_{\omega}\times \nabla' \left( \frac{e^{i\vec{k}\vec{r}}}{r} \right) \mathrm{d}{\bf V'},
\label{elmagfield:eq}
\end{align}
where $\vec{k}= \frac{\omega}{c}\hat{r}$, $\hat{r}$ is a unit vector along vector $\vec{r}$ from the observer to the particle coordinate, $r=|\vec{r}|$ and the spectral components of the charge and current densities are:
\begin{align}
&\rho_\omega(x_s')=\frac{1}{2\pi} \int \rho(x_s',\tau) e^{i\omega \tau} \mathrm{d}\tau
\\
&\vec{j}_\omega(x_s')=\frac{1}{2\pi} \int \vec{j}(x_s',\tau)e^{i\omega \tau} \mathrm{d}\tau
\label{currentdensityfourier:eq}
\end{align}
The common approach is to drop the static 1/${r^2}$ term when differentiating  eq. \ref{elmagfield:eq}, but we will have to impose an even stricter condition later. After differentiating eq. \ref{elmagfield:eq} we obtain:
\begin{align}
&E_\omega(x_s)=\frac{i}{4\pi \epsilon_0} \int \vec{k}\rho_\omega \frac{e^{i\vec{k}\vec{}r}}{r} \mathrm{d}{\bf V'} + \frac{ik}{4\pi \epsilon_0 c} \int \vec{j}_\omega \frac{e^{i\vec{k}\vec{r}}}{r} \mathrm{d} {\bf V'} \\
&B_\omega(x_s)=\frac{i}{4\pi \epsilon_0 c^2} \int \left[ \vec{j}_\omega \times \vec{k} \right] \frac{e^{i\vec{k}\vec{r}}}{r} \mathrm{d} {\bf V'}\label{bomega:eq}
\end{align}
In the far zone, where $\vec{k}\vec{r} >> 1$, $\vec{E} \perp \vec{B}$, the spectral component of the electric field can be written as: 
\begin{align*}
\vec{E}_\omega=\left[ \vec{B}_\omega \times \hat{r} \right]
\end{align*}
or from eq. \ref{bomega:eq}:
\begin{align}
&E_\omega(x_s)=\frac{i\omega}{4\pi \epsilon_0 c^2} \int \left[ \hat{r}\times \left[ \hat{r}\times\vec{j}_\omega\right] \right] \frac{e^{i\vec{k}\vec{r}}}{r} \mathrm{d} {\bf V'}
\label{efieldspeccomp:eq}
\end{align}
In the Fraunhofer zone $R >> \left| \vec{\chi}(t') \right|$ during the whole observation time, where $R=|\vec{R}|$ and $\vec{R}$ is a vector from the observer to the source coordinate origin. Also $r(t')\approx R-\hat{R} \cdot \vec{\chi}(t')$, where $\hat{R}$ is a unit vector along $\vec{R}$, thus, we can replace $r$ by $R$  and $\hat{r}$ by $ \hat{R}$ in eq. \ref{efieldspeccomp:eq}.
For a charge $q$ moving along the trajectory $\vec{\chi}(t)$, the current density can be expressed through the Dirac delta function: 
\begin{align}
\vec{j}(x'_s,t')=q \int \frac{d\vec{\chi}(\tau)}{d\tau} \delta(\vec{r}(t')-\vec{\chi}(\tau)) \mathrm{d} \tau
\label{chargedesitydirac:eq}
\end{align}
Substituting eq. \ref{currentdensityfourier:eq} and eq. \ref{chargedesitydirac:eq} into eq. \ref{efieldspeccomp:eq} and taking into account that $\vec{\beta}=\frac{1}{c}\frac{d\vec{\chi}(t)}{dt}$ and $\vec{k}=n\frac{\omega}{c}\hat{r}$ one can obtain the expression for the spectral component of the electric field:
\begin{align}
&\vec{E}_\omega(x_s)=\frac{i \omega q}{8 \pi^2 \epsilon_0 c} \frac{e^{ikR}}{R} \int \vec{\beta}_\perp e^{i\omega (t-\frac{n\hat{R} \vec{\chi}(t)}{c})} \mathrm{d}t
\label{efieldspecalmostfinal:eq}
\end{align}
where $\vec{\beta}_\perp=\left[ \hat{r} \times \vec{\beta} \right] \times \hat{r}$ is the perpendicular component of the charge velocity directed along $\hat{R}$. 

\begin{figure}[htp]
\includegraphics[width=0.8\columnwidth]{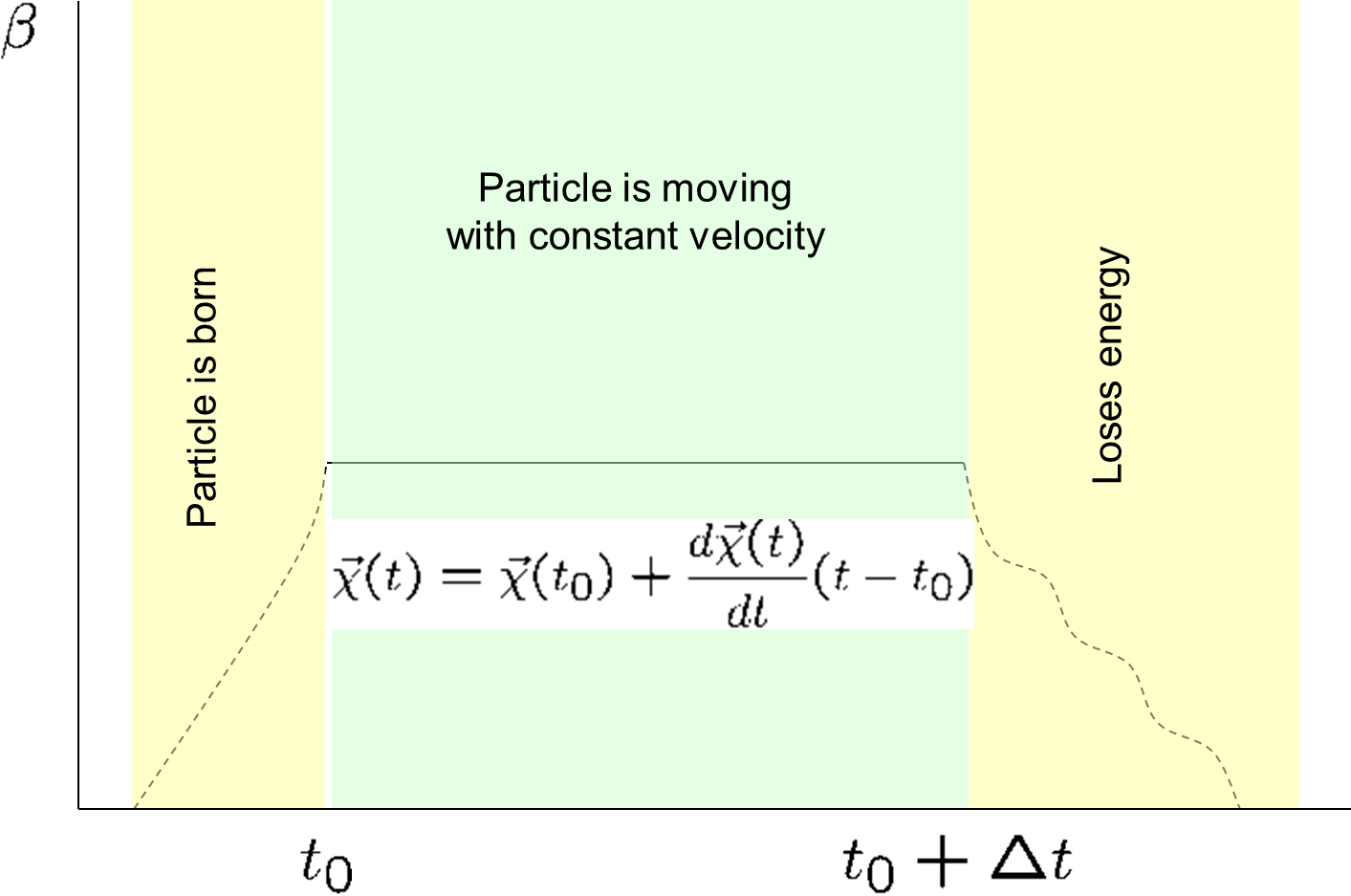}
\caption{Particle velocity while moving along a track segment.}
\label{VelocityDiagram:fig} 
\end{figure}

For the particle acceleration up to time $t_0$, see the diagram on Figure \ref{VelocityDiagram:fig},
the spectral component of the electric field is:
\begin{align}
&\vec{E}_\omega=-i\omega\frac{q e^{ikR} e^{i \omega (t_0 - \frac{1}{c} n \hat{R} \vec{\chi}(t_0))}}{8\pi \epsilon_0 c R}
\int \vec{\beta}_\perp \mathrm{d}t
\label{leftend:eq}
\end{align}
The integral is eq. \ref{leftend:eq} can be approximated by $\beta sin\theta t_0$, where $\theta$ is the angle between $\vec{\beta}$ and $\vec{R}$. Since $\beta$ and $sin\theta$ cannot exceed 1, the whole integral can not exceed $t_0$. If $t_0 << \Delta t$, we can safely neglect the emission from that end of the track segment. Similar conclusions can be made for the deceleration end of the track segment. Integrating the remaining part of the expression \ref{efieldspecalmostfinal:eq} and summing over all track segments we obtain:
\begin{align}
&\vec{E}_\omega(x_s)=\frac{q}{8 \pi^2 \epsilon_0 c} \sum_m \frac{e^{ikR_m}}{R_m} e^{i\omega(t_0^m-\frac{n\hat{R_m}\chi^m(t_0)}{c})} \times \nonumber \\
&\frac{\vec{\beta}_{m \perp}}{(1-n\hat{R}_m\vec{\beta}_m)}\left[ e^{i\omega \Delta t_m (1-n\hat{R}_m\vec{\beta}_m)}-1 \right],
\label{efieldsolution:eq}
\end{align}
where $m$ is the segment index.

\subsection{Summing over the emission from the ends of the track.}

In this approach we account for radiation from the ends of the track $\vec{\chi}(x'_s)$, see the diagram on Figure \ref{ChargedDiagram:fig}, where the charged particle accelerates and decelerates, and ignore the radiation from a straight portion of the track where particle moves without an acceleration, assuming the particle moves in vacuum first and adding the index of refraction later. We start with the equation for the electric field \cite{landaufieldtheory:88}:
\begin{align}
\vec{E}(t,x_s)=&\frac{q}{4\pi \epsilon_0}
\left[ \frac{(\hat{r}-\vec{\beta})(1-\beta^2)}{r^2 (1-\hat{r}\vec{\beta})^3} \right] + \nonumber 
\\
&\frac{q}{4 \pi \epsilon_0 c}
\left[ \frac{\hat{r}\times \left[ (\hat{r}-\vec{\beta}) \times \frac{d \vec{\beta}}{dt} \right] }{r(1-\hat{r}\vec{\beta})^3} \right]
\label{efieldlandau:eq}
\end{align}
where all values are taken at a delayed time $t'$.
In the Fraunhofer zone $\hat{R} \approx \hat{r}$, see Figure \ref{ChargedDiagram:fig}, and magnetic and electric fields are perpendicular:
$\vec{B}(t,x_s)= \frac{1}{c}\vec{E}(t, x_s)\times \hat{r}$. Also note that $\frac{dt}{dt'}=1-\hat{r}\vec{\beta}$. For the spectral component of the electric field we obtain:
\begin{align}
&\vec{E}_\omega (x_s)= \frac{q}{8\pi^2 \epsilon_0 c}
\int \frac{\hat{r} \times \left[ (\hat{r}-\vec{\beta}) \times \frac {d\vec{\beta}}{dt'}\right]}{r(1-\hat{r}\vec{\beta})^2} e^{i \omega (t'+\frac{\vec{\chi}(t')}{c})} \mathrm{d}t'
\end{align}
Taking into account that:
\begin{align}
\frac{d}{dt'} \left[ \frac{\hat{r} \times \left[ \hat{r} \times \vec{\beta} \right]}{1-\hat{r}\vec{\beta}} \right] =
\frac{\hat{r} \times \left[ (\hat{r}-\vec{\beta}) \times \frac{d\vec{\beta}}{dt'} \right] }{(1-\hat{r}\vec{\beta})^2}
\end{align}
we finally obtain for the electric field:
\begin{align}
&\vec{E}_\omega (x_s)= \frac{q}{8 \pi^2 \epsilon_0 c} \frac{e^{ikR}}{R}
\int \frac{d}{dt'} \left[ \frac{ \vec{\beta}_\perp } {1-\hat{R}\vec{\beta}} \right]
e^{i\omega (t'- \frac{\hat{R}\vec{\chi}(t')}{c})} \mathrm{d}t'.
\label{e_omega20:eq}
\end{align}
Solving eq. \ref{e_omega20:eq} for particle birth or acceleration end of the track segment, where $t<t_0$, yields:
\begin{align}
&\vec{E}_\omega (x_s)=-\frac{q}{8 \pi^2 \epsilon_0 c} \frac{e^{ikR}}{R} e^{i\omega (t_0- \frac{\hat{R}\vec{\chi}(t_0)}{c})} \frac{\vec{\beta}_\perp}{1-\hat{R}\vec{\beta}},
\end{align}
and for the deceleration end, where $t>t_0+\Delta t$, yields:
\begin{align}
&\vec{E}_\omega (x_s)=\frac{q}{8 \pi^2 \epsilon_0 c} \frac{e^{ikR}}{R}
e^{i\omega (t_0 + \Delta t - \hat{R}(\frac{\vec{\chi}(t_0)}{c}+ \Delta t \vec{\beta}))}
\frac{\vec{\beta}_\perp}{1-\hat{R}\vec{\beta}}
\end{align}
We can neglect how the particle velocity change during acceleration or deceleration if $\omega \delta t << 1$. Finally, adding the emission from both ends of the track segment, accounting for the index of refraction and summing over all segments of the track we arrive to eq. \ref{efieldsolution:eq} again. Note that the index of refraction was added to the solution for the spectral component of the electric field at the very end. Strictly speaking this is not correct because we started with the full spectrum equation for the electric field \ref{efieldlandau:eq}. In a narrow frequency band however, where the medium dispersion can be neglected, such an approximation is valid.

\section{Coherence wavelength and coherence time.}

\begin{figure}[htp]
\includegraphics[width=\columnwidth]{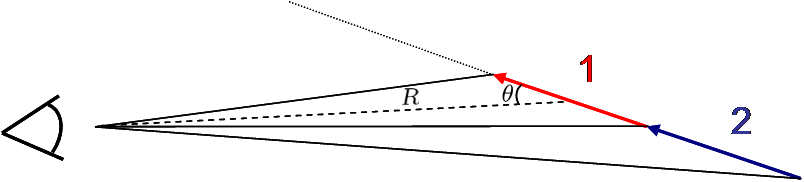}
\caption{Particle trajectory split into Fresnel zones.}
\label{CoherentWavelenght:fig} 
\end{figure}

We split the particle trajectory $\vec{\chi}(x'_s)$ into Fresnel zones so that the phase difference between nearby zones is equal to $\pi$, see the diagram in Figure \ref{CoherentWavelenght:fig},
so that $e^{i \omega t(1-\hat{R}_2\vec{\beta}_2) + i \pi} = e^{i \omega t(1-\hat{R}_1\vec{\beta}_1)}$, where $R_1$ and $R_2$ is the distance from the observer to the corresponding Fresnel zone, and $\vec{\beta}_1$ and $\vec{\beta}_2$ are corresponding velocities. In the Fraunthofer limit $R_1 \approx R_2 \approx R$,  $\beta_1 \approx \beta_2 \approx \beta$ and $\Delta t = t_2-t_1$. Then $\Delta t \omega (1 - \hat{R} \vec{\beta}) = \pi$ or
\begin{align}
c \beta \Delta t = \frac{c \beta \pi}{\omega(1-\hat{R} \vec{\beta})}
\label{cbetadeltat:eq}
\end{align}
note that $\frac{c \pi}{\omega}=\frac{\lambda}{2}$ and $\hat{R}\vec{\beta}=\beta cos\theta$, where $\theta$ is the observation angle. The left hand side in the eq. \ref{cbetadeltat:eq} has the dimension of length, which is called the coherence length:\footnote{Since eq. \ref{lcoh:eq} is not valid at the Cherenkov angle, $\theta_c$, another expression should be used: $L_{coh}=\frac{\sqrt{R\lambda}}{sin\theta_c}$. Related definitions were formulated by I. Tamm in 1939 \cite{tamm:75}. }
\begin{align}
&L_{coh}=\frac{\beta \lambda}{2(1-n\beta cos\theta)}
\label{lcoh:eq}
\end{align}
A corresponding coherence time:
\begin{align}
\Delta t_{coh}=\frac{\pi}{\omega(1-\hat{R}\vec{\beta})}.
\end{align}
The integration along the whole track can be replaced by the integration over one Fresnel zone. For example, using the Fresnel zone closest to the observer for the emission by the particle coming from infinity and stopping at time $t_0$: 
\begin{align}
&\int_{- \infty}^{t_0} \vec{\beta}_\perp e^{i\omega (t-\frac{n\hat{R} \vec{\chi}(t)}{c})} \mathrm{d}t = \frac{1}{2}\int_{t_0-\frac{\pi}{\omega (1-\hat{R_1}\vec{\beta})}}^{t_0} \vec{\beta}_\perp e^{i\omega (t-\frac{n\hat{R_1} \vec{\chi}(t)}{c})}\mathrm{d}t.
\end{align}
The choice of the Fresnel zone is arbitrary. Although the particle is emitting from the whole track, it appears for the observer as if all the emission comes from the track's end. The emission from the nearby Fresnel zones is canceled out due to the phase being different by $\pi$, with exception for the closest, or any other arbitrary chosen Fresnel zone, where an asymmetry is imposed.  One practical application of this is the reduction of the computing time for the MC simulations of the radio emission from air showers.

\section{Conclusions.}

Both approaches presented above are mathematically equivalent if both derived under $\vec{k}\vec{r} >> 1$ condition. In a low frequency limit as well as close to the Cherenkov angle, the second approach can go to infinity, while the first give a finite answer, but both approaches, strictly speaking, are not valid. One also should be careful about subtracting very close numbers doing MC simulation on a computer when using the second approach. As noted before, adding the effects of the medium into the second approach is not straightforward, but a simple addition of the index of refraction is valid as an approximation for some special cases. Another interpretation of the problem is proposed in \cite{konstantinov:phd}. The MC simulation codes described in \cite{AlvarezMuniz2012325} and \cite{Ludwig2011438} are based on the first and second approaches respectively. The results of these MC simulations have never been validated by an accelerator experiment. The full solution without the far zone limitation can be obtained, but the resulting integral contains a modified Bessel function of the second kind and can not be solved analytically.
Finally, it should be noted that we should interpret the emission in the presence of medium as emission from \emph{all} moving  particles, including the charges in the medium. We did not account for the particle interaction with the charges in the medium in both approaches. This is not a problem for the MC simulations on the computer.

%%%%%%%%%%%%%%%%%%%%%%%%%%%%%%%%%%%%%%%%%%%%%%%%
%% BACKMATTER
%%%%%%%%%%%%%%%%%%%%%%%%%%%%%%%%%%%%%%%%%%%%%%%%

% \begin{theacknowledgments}

% \end{theacknowledgments}

%%%%%%%%%%%%%%%%%%%%%%%%%%%%%%%%%%%%%%%%%%%%%%%%
%% The bibliography can be prepared using the BibTeX program or
%% manually.
%%
%% The code below assumes that BibTeX is used.  If the bibliography is
%% produced without BibTeX comment out the following lines and see the
%% aipguide.pdf for further information.
%%
%% For your convenience a manually coded example is appended
%% after the \end{document}
%%%%%%%%%%%%%%%%%%%%%%%%%%%%%%%%%%%%%%%%%%%%%%%%

%%%%%%%%%%%%%%%%%%%%%%%%%%%%%%%%%%%%%%%%%%%%%%%%
%% You may have to change the BibTeX style below, depending on your
%% setup or preferences.
%%
%%
%% For The AIP proceedings layouts use either
%%%%%%%%%%%%%%%%%%%%%%%%%%%%%%%%%%%%%%%%%%%%

\bibliographystyle{aipproc}   % if natbib is available
%\bibliographystyle{aipprocl} % if natbib is missing

%%%%%%%%%%%%%%%%%%%%%%%%%%%%%%%%%%%%%%%%%%%
%% You probably want to use your own bibtex database here
%%%%%%%%%%%%%%%%%%%%%%%%%%%%%%%%%%%%%%%%%%%
\bibliography{KBelovRFAirShowers}
%\bibliography{sample}
\end{document}